\documentclass[a4paper]{amsart}
\usepackage[T1]{fontenc}
\usepackage[utf8]{inputenc}
\usepackage[italian, english]{babel}

\usepackage{amsmath}
\usepackage{amssymb}
\usepackage{amsthm}
\usepackage{hyperref}
\usepackage{subfig}
\usepackage{graphicx}
\usepackage{amsaddr}
\usepackage{xcolor}
\hypersetup{
    colorlinks,
    linkcolor={red!50!black},
    citecolor={blue!50!black},
    urlcolor={blue!80!black}
}

\author{D.~Riccobelli}
\address{SISSA -- International School for Advanced Studies,\\
via Bonomea 265, 34136 Trieste, Italy.}
\email{davide.riccobelli@sissa.it}
\author{D. Ambrosi}
\address{DISMA, Politecnico di Torino,\\
corso Duca degli Abruzzi 24, 10129 Torino, Italy}
\email{davide.ambrosi@polito.it}
\theoremstyle{plain} 

\DeclareMathOperator{\Grad}{Grad}

\DeclareMathOperator{\tr}{tr}

\usepackage{mathtools}
\usepackage{pgfplots}
\pgfplotsset{/pgf/number format/use comma,compat=newest}

\renewcommand\L{\mathcal{L}}

\newcommand{\R}{\mathbb{R}}

\newcommand{\vect}[1]{\boldsymbol{#1}}
\newcommand{\tens}[1]{\mathsf{#1}}

\begin{document}
\title[Activation as a mapping of stress-strain curves]{Activation of a muscle as a mapping of stress-strain curves}

\begin{abstract}
The mathematical modeling of the contraction of a muscle is a crucial problem in biomechanics. Several different models of muscle activation exist in literature. A possible approach to contractility is the so-called \emph{active strain}: it is based on a multiplicative decomposition of the deformation gradient into an active contribution, accounting for the muscle activation, and an elastic one, due to the passive deformation of the body. 

We show that the active strain approach does not allow to recover the experimental stress-stretch curve corresponding to a uniaxial deformation of a skeletal muscle, whatever the functional form of the strain energy. To overcome such difficulty, we introduce an alternative model, that we call \emph{mixture active strain} approach, where the muscle is composed of two different solid phases and only one of them actively contributes to the active behavior of the muscle.

\end{abstract}
\keywords{Muscle contraction -- Active strain -- Hyperelasticity -- Rank-one convexity}
\maketitle

\section{Introduction}

Many soft materials undergo morphological transitions when they are subjected to external stimuli of non-mechanical (e.g. electrical, chemical) nature. Such materials are called \emph{active} because of their ability to undergo motion even in absence of external forces.

An important example of active material is provided by the muscle tissue, which can contract in presence of an electrical stimulus. A correct constitutive modeling of both the active and passive behavior is crucial for several biomechanical systems, such as the modelling of the heart and of the skeletal muscles. 

A mathematical description of the muscle tissue poses several challenges. First, nonlinear constitutive laws are required since a muscle can undergo large deformations. Moreover, a muscle, seen as a material, is strongly anisotropic due to the presence of muscle fibers; in particular it can be suitably represented as transversely isotropic (as in the case of the skeletal muscle) or orthotropic material (as happens in the myocardium, due to the  occurrence of two families of fibers). Furthermore, the process of activation of a muscle is very complex and involves several mechanisms at the microstructural level \cite{caruel2018physics, regazzoni2018active}.
 
A robust constitutive model accounting for the ability of a muscle to contract is far from being established. During the last few decades, several methods have been developed to model the active behavior of muscles in the framework of continuum mechanics \cite{nardinocchi2007active,ambrosi2012active}. The most popular one is the so called active stress \cite{ambrosi2012active}. Such an approach involves an additive split of the total stress into a passive and an active component \cite{smith2004multiscale,panfilov2005self, niederer2008improved}. Another approach is the \emph{active strain}, a technique based on the theory of elastic distortions. In a biomechanical context, such an approach was first introduced by Kondaurov and Nikitin, and further developed by Taber and Perucchio \cite{kondaurov1987finite, taber2000modeling, nardinocchi2007active}. The active strain has been employed in several models (e.g. \cite{cherubini2008electromechanical, rossi2012orthotropic, nobile2012active, pezzuto2014orthotropic}) due to its robust mathematical properties and the clear physical interpretation: the muscle contraction corresponds to a geometrical remodelling of the microstructure of the body \cite{epstein2015mathematical, riccobelli2018existence}.

While the active strain approach guarantees some suitable mathematical properties, the flexibility of the active stress in general allows a better accordance with the experimental results \cite{ambrosi2012active,rossi2012orthotropic,heidlauf2013treatment,giantesio2017comparison}.
Nevertheless, the preservation of the well posedness of the mathematical problem is not always guaranteed by the latter approach \cite{pathmanathan2010cardiac, ambrosi2012active} and could manifest itself in unexpected numerical issues. The active stress formulation is so general that includes the active strain as a particular case \cite{giantesio2017comparison}.

The aim of this article is to compare experimental data on the uniaxial isometric activation of a skeletal muscle \cite{wilkie1956mechanical,hawkins1994comprehensive} with a stress field predicted by the active strain theory.
We show that a plain active strain approach is intrinsically unable to reproduce experimental data, but suitable modifications can be effective in this respect.

The work is organized as follows: in Section~\ref{sec:rev_act_strain} we review the active strain approach and its mathematical properties, in Section~\ref{sec:mapping} we compare the experimental data of Hawkins and Bey \cite{hawkins1994comprehensive} with the stress-stretch curves predicted by the active strain approach. In Section~\ref{sec:decoupled_act_strain} we propose an alternative model based on the mixture active strain method. 

\section{The active strain approach}
\label{sec:rev_act_strain}

We denote by $\L^+(\R^3)$ the set of all the linear maps $\tens{L}:\R^3\rightarrow\R^3$ with positive determinant. Moreover, we indicate with $\mathcal{U}^+(\R^3)$ the subset of $\L^+(\R^3)$ composed of all the linear applications $\tens{L}$ such that $\det\tens{L} = 1$

Let $\Omega_0$ and $\Omega_\text{e}$ be the reference and the actual configuration of an elastic body respectively. We denote with $\vect{X}\in\Omega_0$ the material position vector and with \mbox{$\vect{\varphi}:\Omega_0\rightarrow\Omega_\text{e}$} the motion function. We denote by $\tens{F}=\Grad\vect{\varphi}$ the deformation gradient tensor.

We assume that the material is incompressible and hyperelastic and we denote by $\psi_0$ the strain energy density of the passive material. Thus, the first Piola--Kirchhoff stress of the passive material is given by
\begin{equation}
\label{eq:Pnonact}
\tens{P}_0=\frac{\partial\psi_0}{\partial\tens{F}}-p\tens{F}^{-T},\qquad (P_0)_{ij}=\frac{\partial\psi_0}{\partial F_{ij}}-p F^{-1}_{ji};
\end{equation}
where $p$ is the Lagrangian multiplier that enforces the incompressibility constraint $\det\tens{F}=1$.

When the body is activated, we assume that the deformation gradient admits a multiplicative decomposition of the form
\begin{equation}
\label{eq:Fdeco}
\tens{F} = \tens{F}_\text{e}\tens{F}_\text{a},
\end{equation}
where $\tens{F}_\text{a}$ accounts for the local distortion of the material due to the activation. 

Such an approach is inspired by the theory of elastoplasticity: the decomposition \eqref{eq:Fdeco} is usually referred as \emph{Kr\"oner-Lee} decomposition. To the best of our knowledge, its first application in the field of biomechanics is due to Kondaurov and Nikitin and it has been perfected by Taber and Perucchio \cite{kondaurov1987finite, taber2000modeling}.

The distortion field $\tens{F}_\text{a}:\Omega_0\rightarrow\mathcal{U}^+(\R^3)$ is to be constitutively prescribed. As far as it concerns the activation of a muscle, we assume that $\tens{F}_\text{a}$ determines no variation in the local volume, hence $\det\tens{F}_\text{a} = 1$.

The activation-induced distortion of the body can lead to a geometrically incompatible configuration, namely there may not exist a vector map such that $\tens{F}_\text{a}$ is its gradient: $\tens{F}_\text{a}$ is not integrable. The integrability of $\tens{F}$ is restored by another component $\tens{F}_\text{e}$ that accounts for the elastic distortion of the body \cite{rodriguez1994stress}.

The tensor field $\tens{F}_\text{e}$ describes the elastic ``deformation'' due to the presence of external and internal forces and to the restoration of the geometrical compatibility. Hence, the strain energy of the activated material $\psi:\Omega_0\rightarrow\Omega_\text{e}$ is given by
\begin{equation}
\label{eq:active_strain}
\psi(\tens{F}) = \psi_0(\tens{F}_\text{e})=\psi_0(\tens{F}\tens{F}_\text{a}^{-1}),
\end{equation}
and the first Piola--Kirchhoff stress tensor reads
\[
\tens{P}=\frac{\partial\psi}{\partial\tens{F}}-p\tens{F}^{-T}=\frac{\partial\psi}{\partial\tens{F}_\text{e}}\tens{F}_\text{a}^{-T}-p\tens{F}^{-T}.
\]

The active strain approach possesses nice mathematical properties. In fact, if the strain energy density $\psi_0$ is rank-one convex or polyconvex, then $\psi$ preserves such properties \cite{neff2003some, ambrosi2012active}.

\section{Activation as a linear mapping}
\label{sec:mapping}

The aim of this section is to compare the experimental results on the isometric uniaxial activation of a skeletal muscle with the stress fields predicted by the active strain approach. We focus on the work of Hawkins and Bey \cite{hawkins1994comprehensive} who performed traction experiments on a rat tibialis anterior muscle in isometric conditions.

We denote the local direction of the fibers by the vector field $\vect{M}$ with $\left|\vect{M}\right|=1$. A common choice for $\tens{F}_\text{a}$, inspired by the microstructural architecture, is given by
\begin{equation}
\label{eq:Gactivestrain}
\tens{F}_\text{a}=(1-\gamma)\tens{M}+\frac{1}{\sqrt{1-\gamma}}(\tens{I}-\tens{M}),
\end{equation}
where $0\leq\gamma<1$ is a parameter that describes the microstructural degree of contraction of the muscle ($0$ corresponds to the relaxed muscle), $\tens{M}=\vect{M}\otimes\vect{M}$ and $\otimes$ denotes the diadic product. By performing such a choice for the active strain $\tens{F}_\text{a}$ we assume that the contraction of the sarcomere preserves the cylindrical symmetry along the axis identified by the direction $\vect{M}$. 

Let us denote with $\tens{F}_\lambda$ the deformation gradient that corresponds to the uniaxial deformation along the anisotropic direction $\vect{M}$, i.e.
\begin{equation}
\label{eq:flamb}
\tens{F}_\lambda=\lambda \tens{M}+\frac{1}{\sqrt{\lambda}}(\tens{I}-\tens{M}).
\end{equation}
Making use of \eqref{eq:Pnonact}, we can define the function $\phi$ as
\[
\phi\left(\lambda\right):=\psi_0(\tens{F}_\lambda),
\]
in the passive case. Because of the specific form of the active deformation \eqref{eq:Gactivestrain}, in a uniaxial deformation \eqref{eq:flamb} the strain energy density takes the specific form
\begin{equation}
\label{eq:phi}
\psi(\tens{F}_\lambda) = \phi\left(\frac{\lambda}{1-\gamma}\right).
\end{equation}

Differentiating $\psi(\tens{F}_\lambda)$ with respect to $\lambda$, we obtain the principal stress in the direction $\vect{M}$:
\begin{equation}
\label{eq:PstrainM}
P_{\vect{M}}(\lambda,\,\gamma) =\frac{d\psi(\tens{F}_\lambda)}{d\lambda},
\end{equation}
so that, exploiting the relation \eqref{eq:phi} and applying the chain rule, we get
\[
P_{\vect{M}}(\lambda,\,\gamma) =\frac{1}{1-\gamma}\phi'\left(\frac{\lambda}{1-\gamma}\right).
\]

Out of a rescaling of the strain and stress, the stress-stretch relation $P(\lambda,\,\gamma)$ is therefore completely characterized by its passive behaviour. In fact, if we know the passive response 
\[
P_{\vect{M}}(\lambda,\,0) =\frac{d\psi_0(\tens{F}_\lambda)}{d\lambda}= \phi'(\lambda),
\]
then if the muscle is activated we can obtain $P_{\vect{M}}(\lambda,\,\gamma)$ by rescaling the variable of the function $\phi(\lambda)$.

Indeed, from \eqref{eq:PstrainM} we can observe that
\begin{equation}
\label{eq:PMactstrain}
P_{\vect{M}}(\lambda,\,\gamma)=\frac{1}{1-\gamma}\phi'\left(\frac{\lambda}{1-\gamma}\right)=\frac{1}{1-\gamma}P_{\vect{M}}\left(\frac{\lambda}{1-\gamma},\,0\right).
\end{equation}
Thus, the stress-stretch curve of the activated muscle can be straightforwardly obtained by rescaling the stress and the strain variables by the same factor $(1-\gamma)^{-1}$.

\begin{figure}[t!]
\centering
\includegraphics[width=0.75\textwidth]{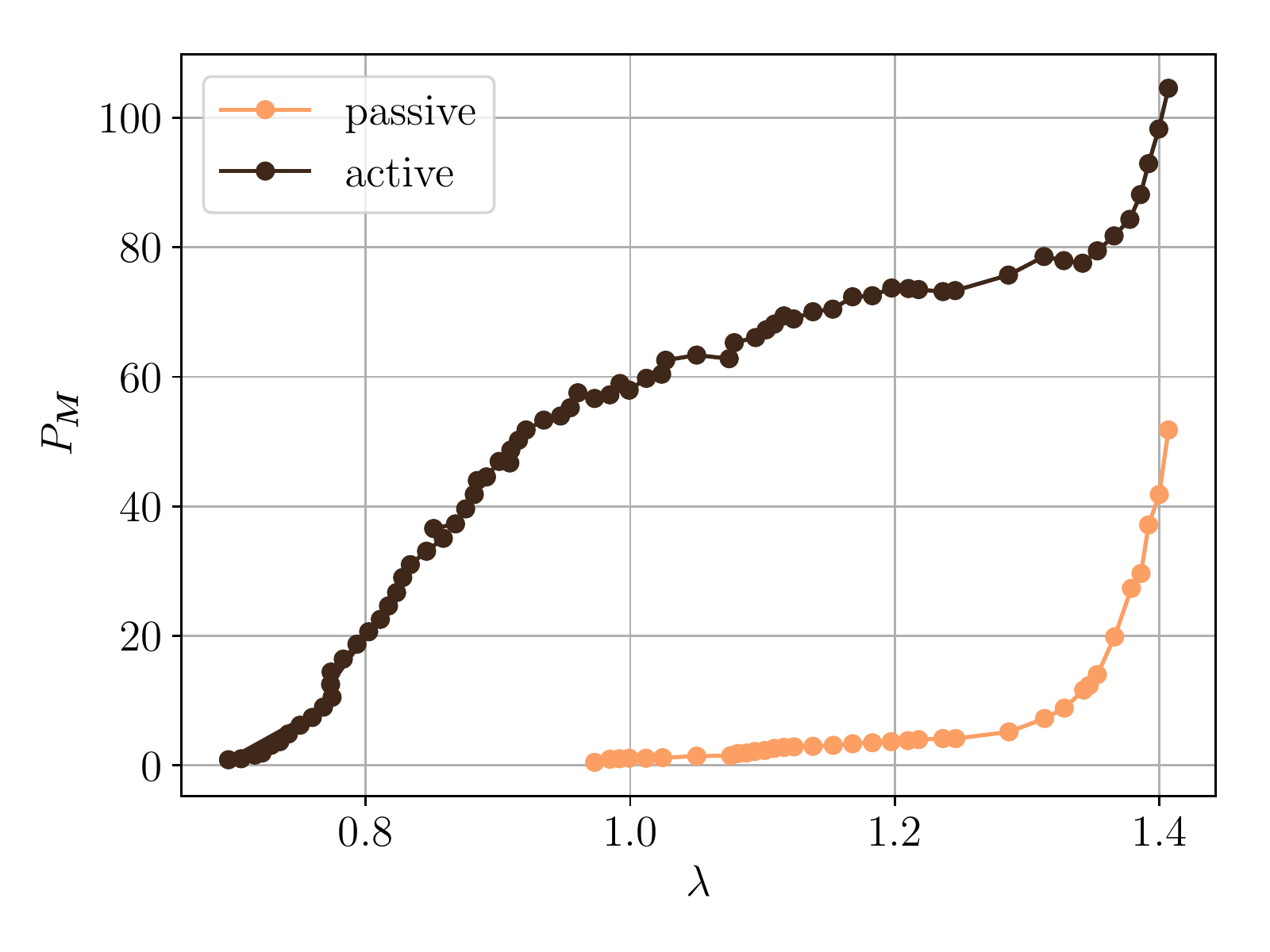}\\\includegraphics[width=0.75\textwidth]{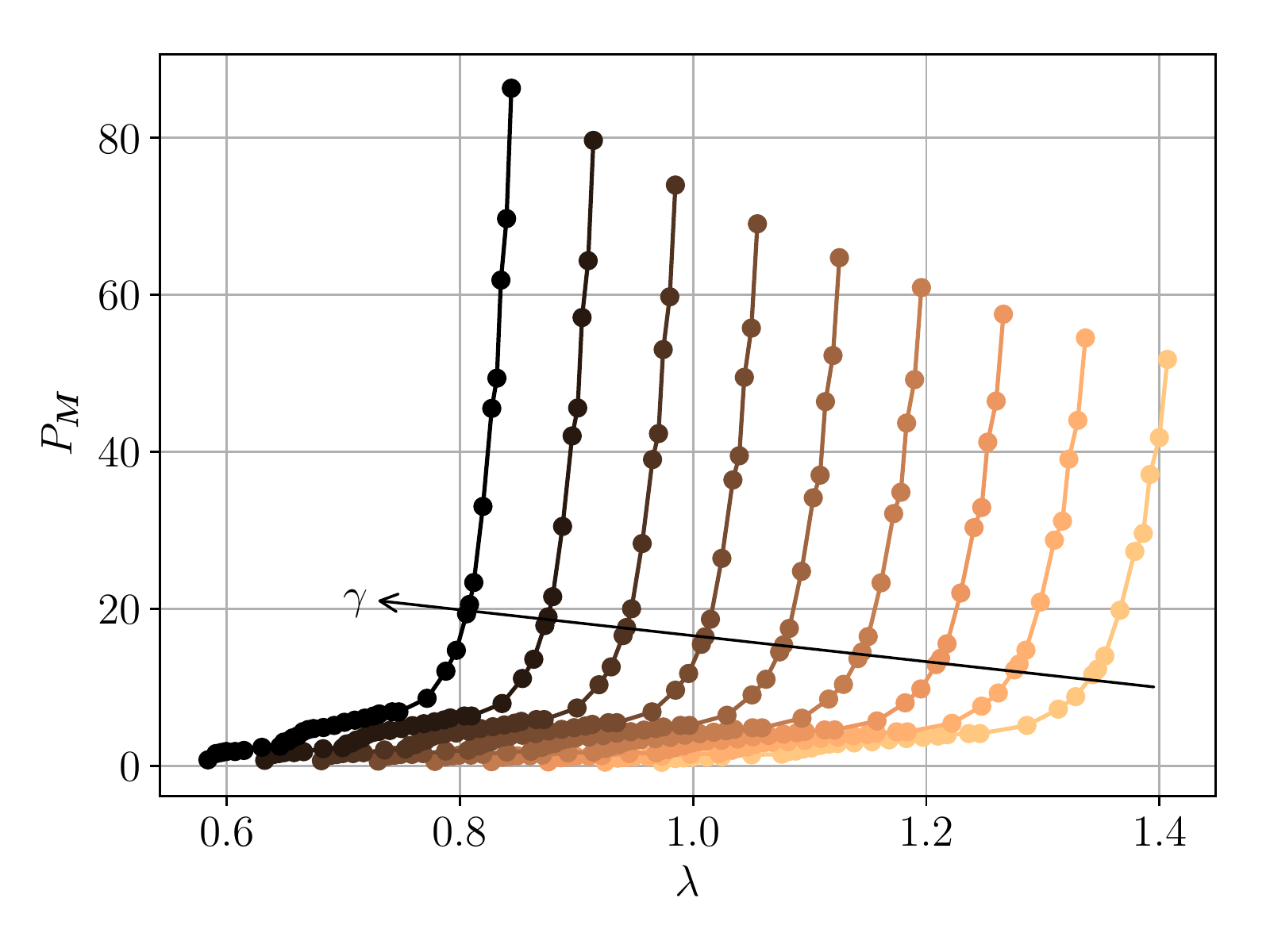}
\caption{(top) Stress-stretch data obtained from the uniaxial traction experiments of Hawkins and Bey \cite{hawkins1994comprehensive}. (bottom) Stress-stretch curves of the uniaxial traction obtained from the passive data by assuming the active strain approach and using \eqref{eq:trans_pass}, here $\gamma$ varies from $0$ up to $0.4$ by steps of $0.05$. The arrow denotes the direction along which $\gamma$ grows.}
\label{fig:dati_gg}
\end{figure}

Unfortunately, such a representation of the activation process is too restrictive to reproduce the experimental results obtained from measuring $P_{\vect{M}}$ in a tetanized muscle \cite{wilkie1956mechanical,hawkins1994comprehensive}. Hawkins and Bey measured the stress-strain relation of the passive muscle and of the tetanized muscle in isometric conditions (Fig.~\ref{fig:dati_gg}). In fact, from the experimental plot in Fig.~\ref{fig:dati_gg} (top) we observe that the passive material exhibits a strain hardening effect when $\lambda\geq1.3$. When the muscle is activated, the curve stress vs strain has a completely different slope: there is a change of concavity and a strain hardening for $\lambda\geq1.3$ and no self-similar transformation reproduces it.

A qualitative attempt to fit of the physiological plot using the active strain approach is obtained by mapping the experimental curve of the passive muscle in Fig.~\ref{fig:dati_gg} (top) into the activated one by the rescaling defined by the equation \eqref{eq:PMactstrain} according to the following procedure. Every point of the stress-stretch plane $(\lambda_0,\,P_0)$ that belongs to the stress-stretch curve of the passive material can be mapped on the activated curve making use of \eqref{eq:PMactstrain}. Indeed, we get that the rescaling
\begin{equation}
\label{eq:trans_pass}
(\lambda,\,P) = \left(\lambda_0(1-\gamma),\,\frac{P_0}{1-\gamma}\right),
\end{equation}
provides the stress-stretch curve of the activated material.

Such a rescaling is applied to the passive curve for several $\gamma$ and the results are shown in Fig.~\ref{fig:dati_gg} (bottom). It is apparent that in this way it is not possible to obtain a stress-stretch curve that fits the experimental data for the contracted muscle. In fact, the strain hardening is anticipated as we increase $\gamma$ and the curve obtained interpolating the experimental data is not convex for $0.7 < \lambda< 1.2$. Thus, the active strain approach cannot reproduce the uniaxial deformation of a contracted skeletal muscle.

We remark that the result of this section do not assume any specific strain energy function to model the passive behavior of the muscle; it is just a rescaling of an experimental curve. Our unique assumption is that the activation of the muscle reads as a contraction along the direction of the fibers that preserves volumes, as in equation \eqref{eq:Gactivestrain}.

\section{An alternative approach: the mixture active strain approach}
\label{sec:decoupled_act_strain}

A possible alternative method is to model the muscle as a material composed of two solid phases, only one of them actively contributing to the muscle contraction.

Let us consider a strain energy density such that
\[
\psi = \psi_\text{iso} + \psi_\text{ani}
\]
where $\psi_\text{iso}$ is the isotropic part of the strain energy whilst $\psi_\text{ani}$ describes the contribution provided by the fibers. We assume that $\psi_\text{iso}$ is only passive and does not give any contribution to the active behavior of the muscle.

The only part of the energy that can provide an active contribution is the function $\psi_\text{ani}$. Thus, we describe the muscle as a mixture of passive (like elastin, randomly distributed collagen) and active materials (like the sarcomeres). We call this approach \emph{mixture active strain} \cite{hernandez20133d,goktepe2014generalized,paetsch2015stability,giantesio2017comparison}.

The first Piola--Kirchhoff stress now reads
\[
\tens{P} = \frac{\partial\psi_{\text{iso}}}{\partial \tens{F}}+\frac{\partial\psi_\text{ani}}{\partial\tens{F}}-p\tens{F}^{-T}
\]
while the principal stress in the direction $\boldsymbol{M}$, denoted by $P_{\boldsymbol{M}}$, produced by the deformation $\tens{F}_\lambda$ is
\[
P_{\vect{M}}(\lambda,\,\gamma)=P_{\vect{M}}^\text{iso}(\lambda)+P_{\vect{M}}^\text{ani}(\lambda,\,\gamma)
\]
where
\[
P_{\vect{M}}^\text{iso}(\lambda) = \frac{d\psi_\text{iso}(\tens{F}_\lambda)}{d\lambda}\qquad P_{\vect{M}}^\text{ani}(\lambda,\,\gamma) = \frac{d\psi_\text{ani}(\tens{F}_\lambda\tens{F}_\text{a}^{-1})}{d\lambda}.
\]
In analogy with \eqref{eq:PMactstrain} we can introduce the following representation
\[
P_{\vect{M}}^\text{ani}(\lambda,\,\gamma) =\frac{1}{1-\gamma}P_{\vect{M}}^\text{ani}\left(\frac{\lambda}{1-\gamma},\,0\right).
\]
Hence, the purely active contribution arising from the contraction of the muscle is due to the anisotropic contribution and it is given by
\[
P^\text{act}_{\vect{M}}(\lambda,\,\gamma)= P_{\vect{M}}(\lambda,\,\gamma)-P_{\vect{M}}(\lambda,\,0)=\frac{1}{1-\gamma}P_{\vect{M}}^\text{ani}\left(\frac{\lambda}{1-\gamma},\,0\right)-P^\text{ani}_{\vect{M}}(\lambda,\,0).
\]
Such a function is expected to fit the experimental data of Hawkins and Bey \cite{hawkins1994comprehensive} relative to the active contribution to the stress $P_{\vect{M}}$ (see Fig.~\ref{fig:dati_gg}). Indeed, setting $\gamma=\gamma_\text{max}$ corresponding to the maximal contraction of the sarcomere, we get that $P^\text{act}_{\vect{M}}(\lambda,\,\gamma_\text{max})$ should reproduce the difference between the stress generated by the tetanized muscle and the stress generated by the passive body: to perform this comparison, we have to chose a specific strain energy density.

Let us introduce the following invariants
\[
I_1 = \tr \tens{C},\qquad J=\det\tens{F},\qquad I_4 = \tr(\tens{C}\tens{M}),
\]
where $\tens{C} = \tens{F}^T\tens{F}$ is the right Cauchy-Green tensor.

We choose to model the isotropic part of the muscle as a Gent material \cite{gent1996new}, so that the strain energy density is given by
\begin{equation}
\label{eq:gent}
\psi_\text{iso}(\tens{F})=-\frac{\mu I_\text{max}}{2}\log\left(1-\frac{I_1 - 3}{I_\text{max}}\right)
\end{equation}
where $\mu$ is the shear modulus and $I_\text{max}$ is a parameter that sets the maximum value reachable by $I_1$.

The anisotropic part of the strain energy is instead given by
\begin{equation}
\label{eq:anisotropic_ener}
\psi_\text{ani}(\tens{F})=\alpha\beta\left({I_4}^{\frac{1}{2\beta}}-1\right)^2.
\end{equation}
Thus the fibres contribute to the strain energy only if there is a deformation in the direction $\vect{M}$. We remark that the constitutive choice \eqref{eq:gent}, while specific, is very popular in the mechanics of soft tissues \cite{gent1996new, horgan2002constitutive, horgan2003description, rashid2012mechanical, rashid2014mechanical}. The anisotropic component of the strain energy \eqref{eq:anisotropic_ener} usually in literature involves the square root $I_4$ \cite{Chagnon_2014}. In this work we adopt a power law with exponent $1/(2\beta)$ to account for the reported change in convexity of $P_{\vect{M}}$ versus $\lambda$ (see Fig.~\ref{fig:dati_gg} top).

The total strain energy of the activated material is hence given by
\begin{equation}
\label{eq:part_act_strain}
\psi(\tens{F}) = \psi_\text{iso}(\tens{F})+(\det{\tens{F}_\text{a}})\psi_\text{ani}(\tens{F}\tens{F}_\text{a}^{-1})
\end{equation}
where $\tens{F}_\text{a}$ is given by \eqref{eq:Gactivestrain}.

\begin{figure}[b!]
\centering
\includegraphics[width = 0.75\textwidth]{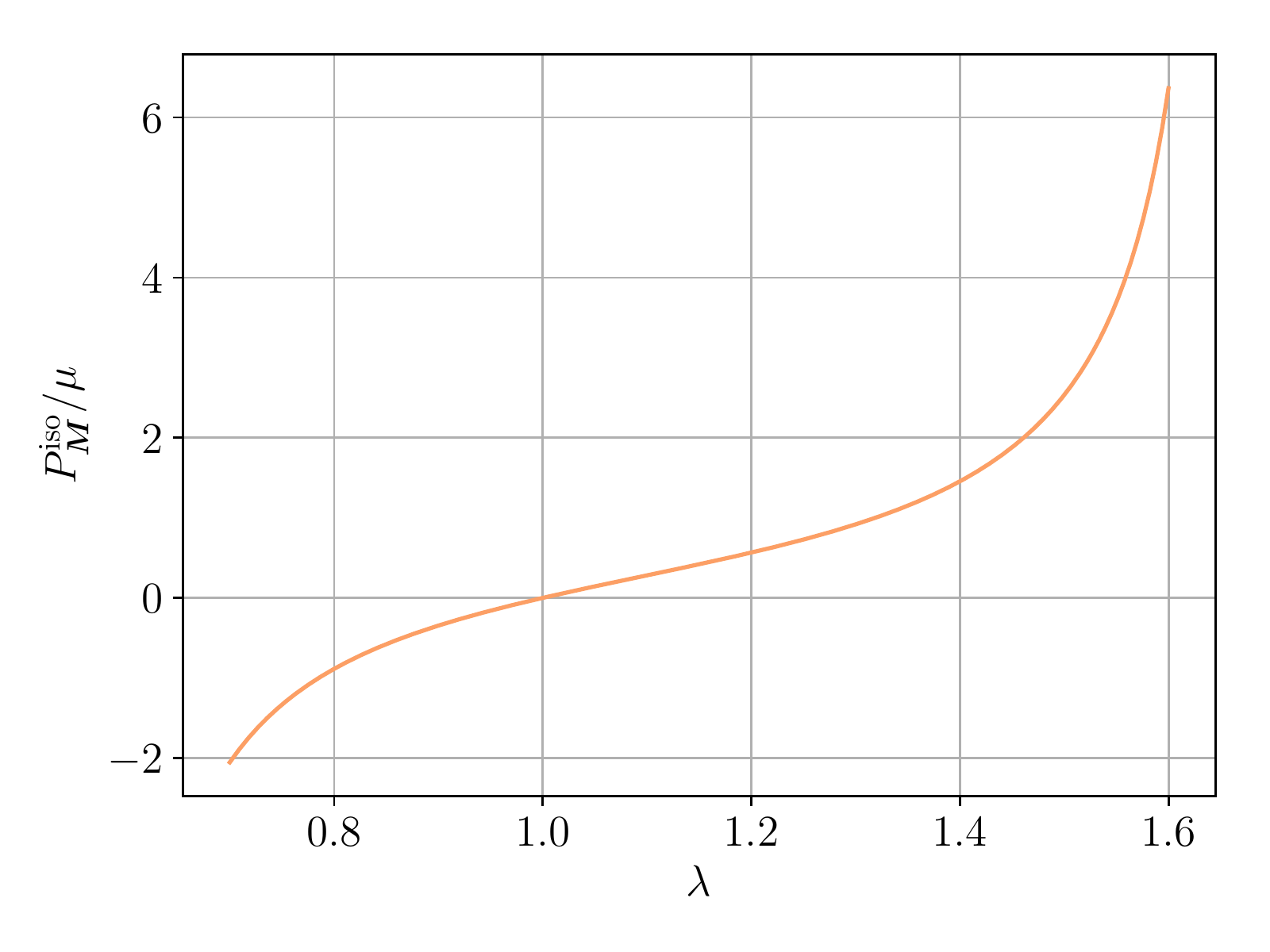}\\\includegraphics[width = 0.75\textwidth]{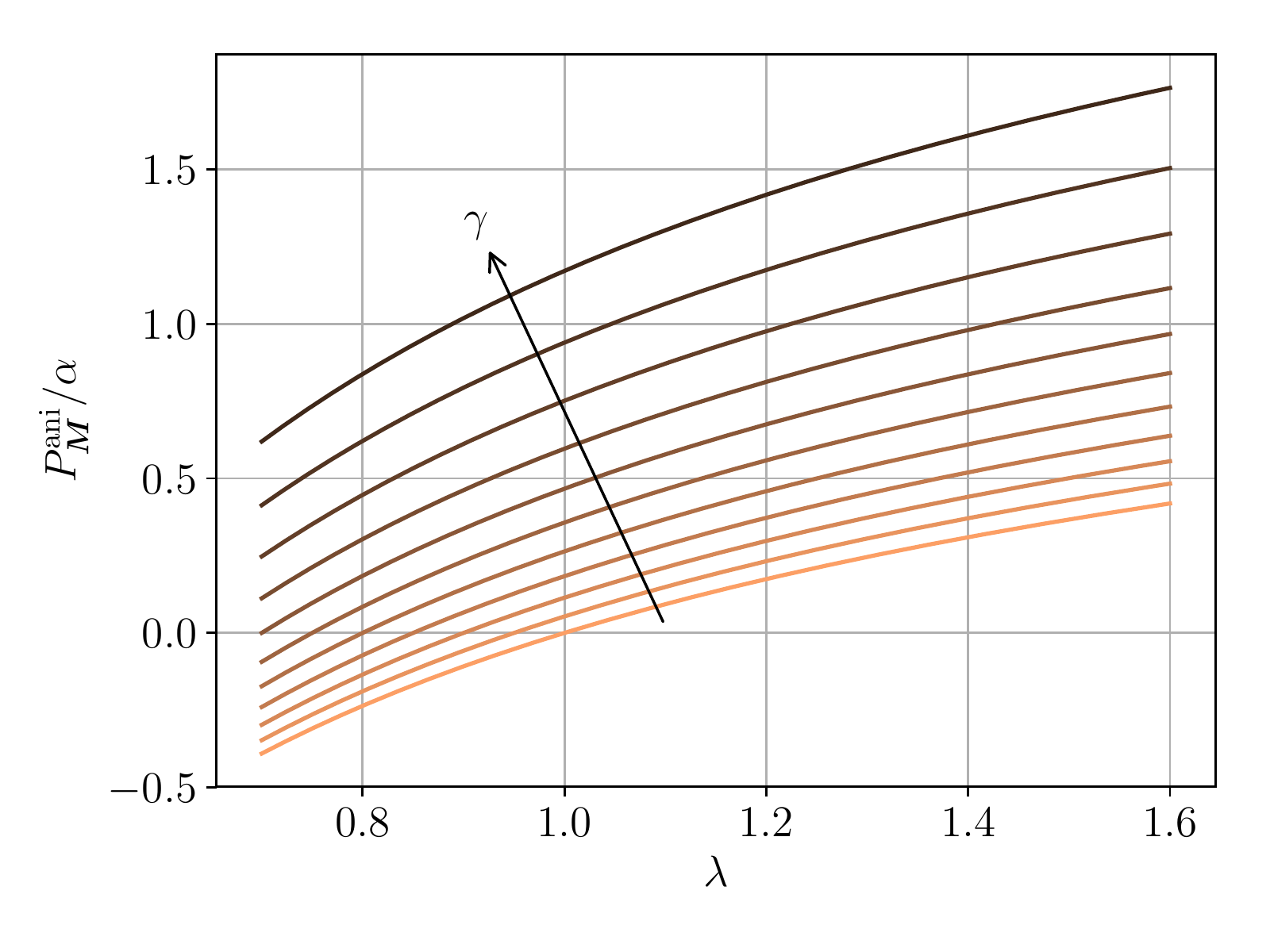}
\caption{Plot of the normalized principal stresses $P^\text{iso}_{\vect{M}}/\mu$ (top) and $P^\text{ani}_{\vect{M}}/\alpha$ (bottom). We set $I_\text{max} = 1$ and $\beta=2$. In the bottom plot, $\gamma$ varies between $0$ and $0.5$ by steps of $0.05$.}
\label{fig:prinStresses}
\end{figure}

Thus, for $\det\tens{F}_\text{a}=1$ the Piola--Kirchhoff stress tensor reads
\[
\tens{P}(\tens{F})=\tens{P}_\text{iso}(\tens{F})+\tens{P}_\text{ani}(\tens{F}\tens{F}_\text{a}^{-1})\tens{F}_\text{a}^{-T}-p\tens{F}^{-T},
\]
where
\begin{gather*}
\tens{P}_\text{iso}(\tens{F}) = \frac{\partial \psi_{\text{iso}}}{\partial\tens{F}}=\mu\left(1-\frac{I_1-3}{I_\text{max}}\right)^{-1}\tens{F},\\
\tens{P}_\text{ani}(\tens{F})= \frac{\partial \psi_{\text{ani}}}{\partial\tens{F}} = 2\alpha\frac{{I_4}^{\frac{1}{2\beta}}-1}{{I_4}^\frac{2\beta-1}{2\beta}}\tens{F}\tens{M}.
\end{gather*}

In particular, in the case of a uniaxial deformation in the direction $\vect{M}$, we can compute the principal stresses $P_{\vect{M}}$ and $P_{\vect{M}}^\text{ani}$ in the direction $\vect{M}$, namely
\begin{gather*}
P_{\vect{M}}^\text{iso}(\lambda) = \frac{d\psi_\text{iso}(\tens{F}_\lambda)}{d\lambda} =\mu\left(1-\frac{\lambda^2+2\lambda^{-1}-3}{I_\text{max}}\right)^{-1}\left(\lambda-\lambda^{-1}\right)\\
P_{\vect{M}}^\text{ani}(\lambda,\,\gamma)= \frac{d\psi_\text{ani}(\tens{F}_\lambda\tens{F}_\text{a}^{-1})}{d\lambda}= \frac{2 \alpha \left(\frac{\lambda}{1-\gamma}\right)^{1/\beta} \left(\left(\frac{\lambda}{1-\gamma}\right)^{1/\beta}-1\right)}{\lambda}.
\end{gather*}

\begin{figure}[t!]
\centering
\includegraphics[width = 0.75\textwidth]{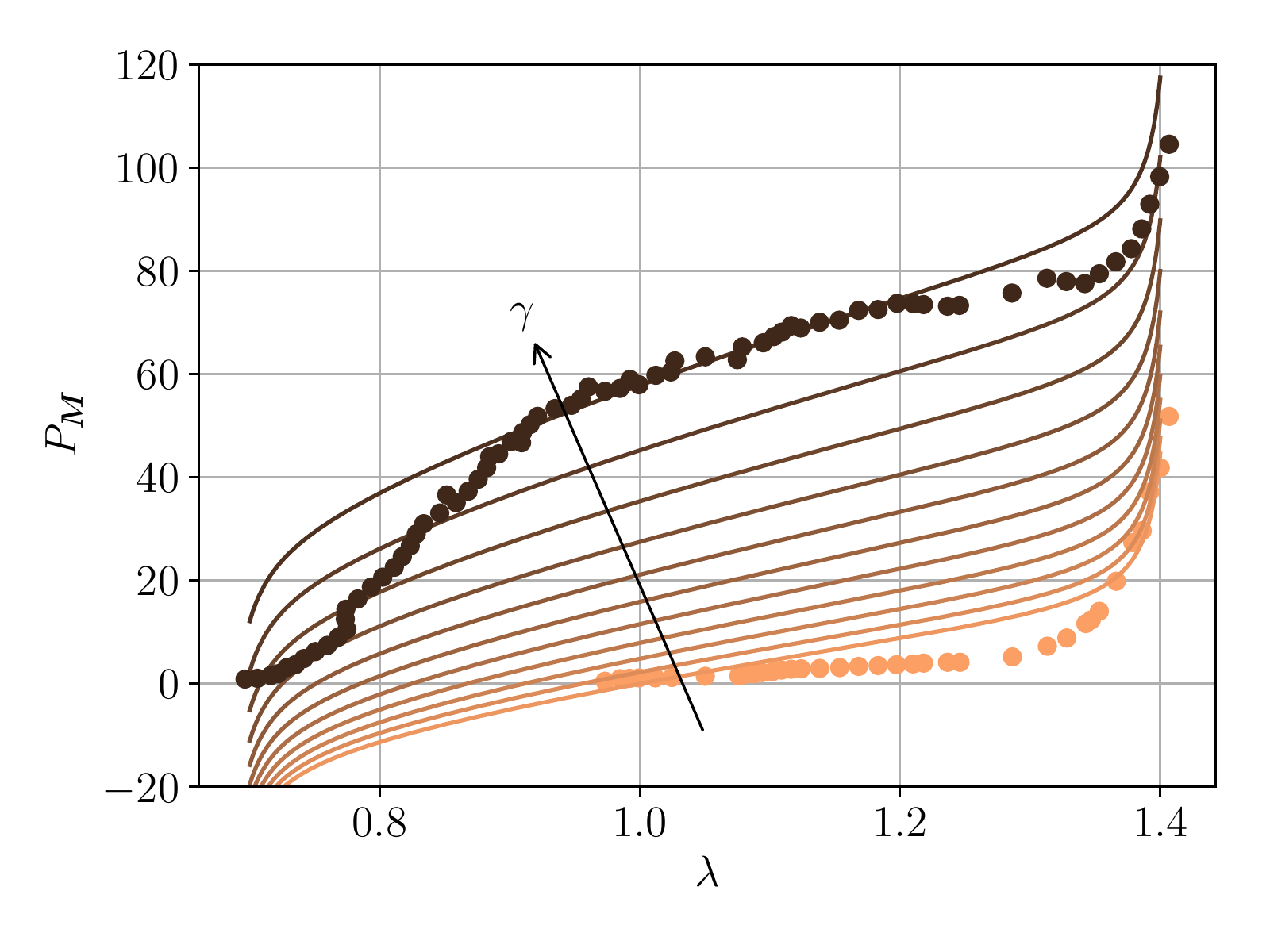}
\caption{Plot of $P_{\boldsymbol{M}}$ when $\mu = 1.8$~kPa, $I_\text{max} = 0.41$, $\alpha = 31$~kPa, $\beta = 1.5$ and $\gamma$ varies from $0$ to $0.5$ with steps of $0.05$.}
\label{fig:ptot}
\end{figure}

Thus, the total principal stress 
\[P_{\vect{M}}(\lambda,\,\gamma)=P_{\vect{M}}^\text{iso}(\lambda)+P_{\vect{M}}^\text{ani}(\lambda,\,\gamma)
\]
 is given by
\[
P_{\vect{M}}(\lambda,\,\gamma) = \mu\left(1-\frac{\lambda^2+2\lambda^{-1}-3}{I_\text{max}}\right)^{-1}\left(\lambda-\lambda^{-1}\right)+\frac{2 \alpha \left(\frac{\lambda}{1-\gamma}\right)^{1/\beta} \left(\left(\frac{\lambda}{1-\gamma}\right)^{1/\beta}-1\right)}{\lambda}.
\]

In Fig.~\ref{fig:prinStresses} we plot the principal stresses $P_{\vect{M}}^\text{iso}(\lambda)$ and $P_{\vect{M}}^\text{ani}(\lambda,\,\gamma)$ normalized with respect to $\mu$ and $\alpha$, respectively. In Fig.~\ref{fig:ptot} we plot the total principal stress $P_{\vect{M}}(\lambda,\,\gamma)$ where we set $\mu = 1.8$~kPa, $I_\text{max} = 0.41$, $\alpha = 31$~kPa and $\beta = 1.5$. Increasing the activation parameter $\gamma$, the stress-stretch relationship is in good agreement with the experimental data of Fig.~\ref{fig:dati_gg} for $\gamma\sim 0.5$.

However, the model has some limitations. The compressive branch of the passive curve can not be compared with the measures of Hawkins and Bey. In the experimental literature, many works report that the passive stress generated performing a uniaxial compression along the direction of the fibers is lower than the one occurring in extension \cite{van2006validated}. Even though other research groups measure stress of the same order of magnitude of the one obtained in extension, in one case even higher \cite{zheng1999objective, bosboom2001numerical, bosboom2001passive} (see \cite{van2006validated} for a comparison between the data), the majority of the experimental works report a softer behaviour in compression, in contrast with our prediction for $\lambda<0.8$ in the passive case. Also the high slope in the active case of the stress-strain curve at the intersection with the $\lambda$-axis is inaccurate. In order to obtain a softer behaviour in compression, it would be necessary to employ a more involved expression of the strain energy densities $\psi_\text{iso}$ and $\psi_\text{ani}$ and it is not the focus of our work (e.g. considering an anisotropic energy also for the passive constituent of the muscle).

Summarizing, we have shown that using an active strain approach for the aniso\-tropic part of the strain energy density only, one can quantitatively reproduce the behaviour of the skeletal muscle in extension. The mixed active strain approach allows to overcome two limitations of the ``pure'' active strain. First, increasing the activation parameter, the strain stiffening appears always at $\lambda=1.3$. Second, we observe a change of concavity in the stress-strain curve in the tetanized case ($\gamma = 0.5$) at $\lambda\simeq1.1$ (Fig.~\ref{fig:ptot}). Mathematical issues raised by the mixture active strain are the subject of the next section.

\subsection{Material symmetry group and muscle activation}

Even if the mixture active strain approach better fits the experimental data with respect to the active strain, it remains an open question whether such approach preserves or not some mathematical properties. For the ``global'' active strain approach \eqref{eq:active_strain}, rank-one convexity is preserved \cite{ambrosi2012active}; in the mixture approach, if the strain energy functional is rank-one convex globally in the passive case, we cannot state rank-one convexity without further assumptions on $\psi_\text{iso}$ and $\psi_\text{ani}$. If both $\psi_\text{iso}$ and $\psi_\text{ani}$ are rank-one convex, then
\[
\delta\tens{F}:\frac{\partial^2\psi(\tens{F})}{\partial\tens{F}\partial\tens{F}}:\delta\tens{F}=\delta\tens{F}:\frac{\partial^2\psi_\text{iso}(\tens{F})}{\partial\tens{F}\partial\tens{F}}:\delta\tens{F}+\delta\tens{F}\tens{F}_a^{-1}:\frac{\partial^2\psi_\text{ani}(\tens{F})}{\partial\tens{F}\partial\tens{F}}:\delta\tens{F}\tens{F}_a^{-1}>0
\]
for all $\delta\tens{F}$ which are rank-one since also $\delta\tens{F}\tens{F}_a^{-1}$ is a rank-one tensor \cite{ambrosi2012active}. If the material is incompressible, $\delta\tens{F}$ must also belong to the tangent space to the manifold $\det\tens{F}=1$, namely $\tens{F}^{-T}:\delta\tens{F}=0$.

The same happens for polyconvexity: if both the $\psi_\text{iso}$ and $\psi_\text{ani}$ are polyconvex in the passive case, then the polyconvexity is preserved in the active case. This is a direct consequence of Lemma~6.5 in \cite{neff2003some}.

The active strain preserves the material properties during muscle contraction. Indeed, the multiplicative decomposition of the deformation gradient is equivalent to a remodelling, leading to a change of the relaxed state of the body \cite{epstein2012elements, epstein2015mathematical}. Other methods, such as the active stress or the mixture active strain, do not correspond to a remodelling and a modification the material properties, such as the shear modulus or the material symmetries, can take place as a consequence of material activation. If we use the language introduced by Epstein \cite{epstein2015mathematical}, there is change of the \emph{archetype} of the body.

It is expected that the symmetry group of the material is preserved, during muscle activation, since the contraction of sarcomeres does not generate any new structural anisotropy. Let 
\begin{gather*}
\mathcal{G}_\text{iso} = \left\{\tens{Q}\in\mathcal{U}^+(\R^3)\;|\;\psi_\text{iso}(\tens{F}\tens{Q})=\psi_\text{iso}(\tens{F})\,\forall\tens{F}\in\mathcal{U}^+(\R^3)\right\},\\
\mathcal{G}_\text{ani} = \left\{\tens{Q}\in\mathcal{U}^+(\R^3)\;|\;\psi_\text{ani}(\tens{F}\tens{Q})=\psi_\text{ani}(\tens{F})\,\forall\tens{F}\in\mathcal{U}^+(\R^3)\right\}.
\end{gather*}
The material symmetry group of the passive muscle is given by $\mathcal{G}=\mathcal{G}_\text{iso}\cap\mathcal{G}_\text{ani}$. If the muscle is activated, the material symmetry group of the anisotropic part of the energy becomes \cite{epstein2015mathematical}
\[
\widehat{\mathcal{G}}_\text{ani}=\tens{F}_\text{a}^{-1}\mathcal{G}\tens{F}_\text{a},
\]
and so the material symmetry group of the whole energy reads $\widehat{\mathcal G}=\mathcal{G}_\text{iso}\cap\widehat{\mathcal{G}}_\text{ani}$.

It is easy to verify that if $\psi_\text{iso}$ is isotropic and $\psi_\text{ani}$ is transversely isotropic  with direction of symmetry $\vect{M}$, if we apply an activation of the form \eqref{eq:Gactivestrain}, then the symmetry group of the material is not modified by the mixture active strain, i.e. $\mathcal{G}=\widehat{\mathcal G}$.

Summarizing, the active strain approach corresponds to a remodelling \cite{epstein2015mathematical, riccobelli2018existence}: there is only a morphological change in the relaxed configuration and the properties of the material are conserved.
In the previous sections, we have proved that the active strain cannot fit the experimental data of the uniaxial extension of a skeletal muscle. During muscle contraction, the mutual positions of the actin and myosin filaments change, leading to a modification of the microstructure: the increased number of cross bridges should result into a different stiffness of the tissue. The active strain approach allows to describe the modification of shape induced by muscle contraction but it does not take into account the evolution of the material properties induced by the formation of cross bridges (i.e. the shear modulus of the material does not change). 

Conversely, the mixture active strain approach correctly reproduces the experimental data for the uniaxial traction of the muscle, without changing the symmetry group of the material.

\section{Discussion of the results and concluding remarks}

In this paper, we have analyzed some aspects related to the modelization of  muscle activation. First, in Section \ref{sec:rev_act_strain} we have provided a review of the active strain approach for modeling the activation of an elastic medium. In Section \ref{sec:mapping} we have compared theoretical predictions vs. experimental data provided by Hawkins and Bey \cite{hawkins1994comprehensive}. We have showed that, independently from the chosen passive model, the active strain approach cannot reproduce the stress-stretch curve of the tetanized tibialis anterior muscle. 

According to the classification made by Epstein \cite{epstein2015mathematical}, the active strain approach corresponds to a remodelling of a material, namely a change of shape that does not affect the material properties and the microstructure. The inadequacy discussed above shows that the contraction of the muscle is not a simple remodelling and the microstructure of the tissue and the material properties change. 

Since it is not possible to model the skeletal muscle contraction as a pure remodelling, in Section \ref{sec:decoupled_act_strain} we have proposed a model of the muscle alternative to the active strain obtained by a mixture approach, applying the Kr\"oner--Lee decomposition of the deformation gradient only on one component (the anisotropic part) of the strain energy density. To make quantitative comparisons, we have used a Gent strain energy density for the isotropic part and the strain energy \eqref{eq:anisotropic_ener} for the anisotropic one. Such a simple approach, called mixture active strain, provides results which are in good agreement with the experimental ones in extension.

Convexity properties are preserved if both the isotropic and the anisotropic part of the strain energy are polyconvex or rank-one convex. Also the material symmetry group is preserved if $\psi_\text{ani}$ is transversely isotropic along the direction $\vect{M}$ and $\tens{F}_\text{a}$ has the form \eqref{eq:Gactivestrain}.

It is to be remarked that while a correct representation of the stretch-stress curve for uniaxial homogeneous deformation is a mandatory requirement, it is not sufficient to obtain a reliable model for a generic deformation, in particular in shear \cite{giantesio2017comparison}; a deeper understanding of the possible change in the microstructure and in the material properties due to the process of muscle activation is required.

The results of this work may support the development of models of  the muscle tissue activity: a reliable mathematical description of the skeletal muscles or of the whole heart are active and open research topics in biomechanics and in the field of biomedical engineering. 

\section*{Acknowledgements}

We are thankful to Giulia Giantesio for providing the data of Fig.~\ref{fig:dati_gg} and Fig.~\ref{fig:ptot} obtained from \cite{hawkins1994comprehensive}.\\
This work has been partially supported by the National Group of Mathematical Physics (\emph{GNFM}--\emph{INdAM}) through the program \emph{Progetto Giovani 2017}.
DA acknowledges the MIUR grant {\em Dipartimenti di Eccellenza 2018-2022}\\ (E11G18000350001).

\section*{Conflicts of interest} All authors declare that they have no conflict of
interest.

\bibliographystyle{abbrv}
\bibliography{refs}
\end{document}